\documentclass{mem}
\usepackage{natbib}\usepackage{txfonts}\usepackage{balance}
\usepackage{flushend}
\usepackage{graphicx}
\usepackage[breaklinks,pdftex]{hyperref}
\usepackage{xcolor}
\usepackage{xspace}
\usepackage{float}
\idline{1}{1}
\begin{document}
\def\teff{$T\rm_{eff }$}
\def\kms{$\mathrm {km s}^{-1}$}
\newcommand{\lat}{{\it Fermi}-LAT\xspace}
\newcommand{\gm}{$\gamma$\xspace}
\newcommand{\pks}{PKS\,1424$-$41\xspace}
\newcommand{\mkn}{Mkn\,421\xspace}

\title{
Recurrence plot analysis of blazar gamma-ray light curves
}

   \subtitle{Exploiting the time-domain capabilities of \textit{Fermi}-LAT}

\author{
Andrea \,Gokus\inst{1} 
\and Rebecca\, Phillipson\inst{2}
}

\institute{
Department of Physics \& McDonnell Center for the Space Sciences, Washington University in St. Louis, One Brookings Drive, St. Louis, MO 63130, USA;
\email{gokus@wustl.edu}
\and
Department of Physics, Villanova University, 800 E Lancaster Ave, Villanova, PA 19085, USA;
\email{rebecca.phillipson@villanova.edu}\\
}

\authorrunning{Gokus \& Phillipson}

\titlerunning{Recurrence plot analysis of blazar \gm-ray light curves}

\date{Received: XX-XX-XXXX (Day-Month-Year); Accepted: XX-XX-XXXX (Day-Month-Year)}

\abstract{
Variability studies of jetted AGN, in particular blazars, have been used to gain a better understanding of the particle acceleration mechanisms in jets. However, statistical methods used for the characterization of variability often rely on stationary time series data, which is not fulfilled for most blazar light curves.
We introduce the recurrence plot method for long-term \gm-ray light curves sampled by \lat and present our results for the BL Lac object \mkn and the FSRQ \pks. 
Using surrogates to determine the significance of our findings, we conclude that \mkn exhibits more determinism than \pks, and that both sources potentially show nonlinearity. However, the latter has to be tested against more advanced surrogates that are able to replicate the nonstationarity of the original light curves. In future work, we will extend our recurrence analysis to a sample of $\sim50$ \gm-ray bright sources to probe the jet dynamics of different blazar classes.
}
\maketitle{}

\section{Introduction}
Blazars belong to the class of jetted active galactic nuclei (AGN). Their jets point toward our line of sight from a small angle, which results in their emission appearing very luminous and variable due to relativistic beaming \citep{urry1995}. Blazars can be divided into two main categories: flat-spectrum radio quasars (FSRQs), for which the optical continuum emission contains emission lines with a width of $\geq5\AA$, and BL Lacertae objects (BL Lacs) that only exhibit narrower lines or none at all.

We can study the particle acceleration processes in the jet utilizing blazar variability. Since the detection of the strong and rapid variations exhibited by blazars, their behaviour has been studied across the electromagnetic spectrum by a multitude of instruments \cite[e.g.,][]{jorstad2001,Bhatta2018,Rajput2020,Bhatta2021}.
However, continuous and unbiased all-sky monitoring is necessary to study the variability of the blazar population as a whole, which is only possible with a few instruments. At \gm-ray energies, the \textit{Fermi} Large Area Telescope \cite[LAT;][]{lat_instrument} has been providing these data since its launch in 2008 for $>3800$ blazars \citep{4lac_dr3}. For the brightest sources, we can use \lat data for studies relying on long-term, high-cadence monitoring with day-to-week time resolution.
A variety of analysis methods that have been applied to blazar light curves across the electromagnetic spectrum exists, such as fractional variability \cite[e.g.,][]{Schleicher2019_fractionalvariability}, periodograms \cite[e.g.,][]{ciaramella2004}, power spectral densities \cite[PSDs; e.g.,][]{abdo2010,ryan2019}, and wavelet analysis \cite[e.g.,][]{humrickhouse2008,ege2024}. 
In recent years, the long-term sampling of data has enabled searching for quasi-periodic oscillations (QPOs), and for several sources, tentative results of multi-year periodicities have been reported \cite[e.g.,][]{oterosantos2023,chen2024,rico2024}. Precessing jets and supermassive black hole binaries have been proposed as possible origins of QPOs \citep{rieger2004}.

Most time series analysis methods rely on the underlying assumption that the data is stationary, with mean and variance remaining constant over time. 
However, this is not the case for blazar light curves, which exhibit rapid flares of varying length.
In this work, we present a different approach to characterize the flux variability observed in bright \gm-ray blazars using the recurrence plot (RP) analysis technique. 
RPs are ideally suited to study blazar light curves as they can retrieve the same information embedded in PSDs without making assumptions about stationarity, probe non-linearity, and provide distinctions between stochastic and deterministic processes.

We introduce the concept of RP analysis in Sect.~\ref{sec:methods}, and present our results based on \lat light curves of the blazars \mkn and \pks in Sect.~\ref{sec:results}.

\begin{figure*}
\centering
\includegraphics[width=0.47\textwidth]{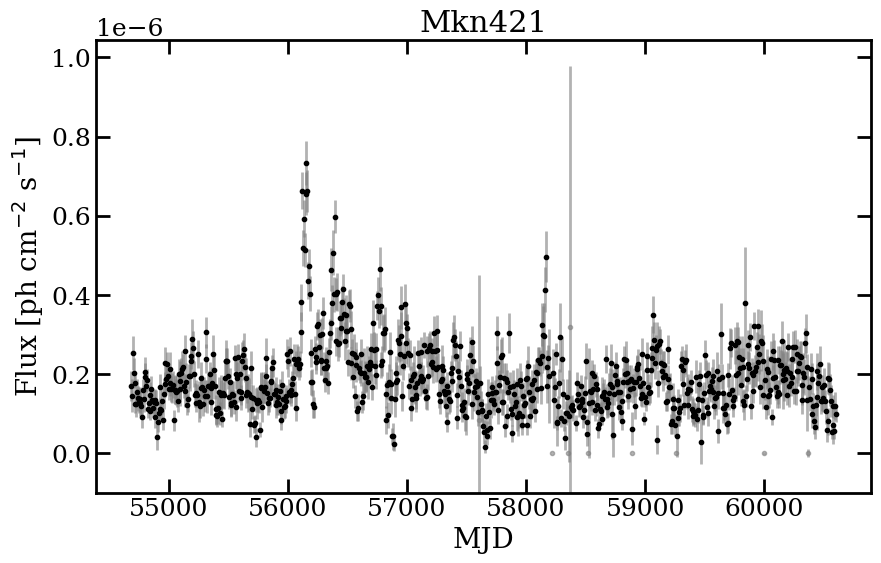}\hfill
\includegraphics[width=0.47\textwidth]{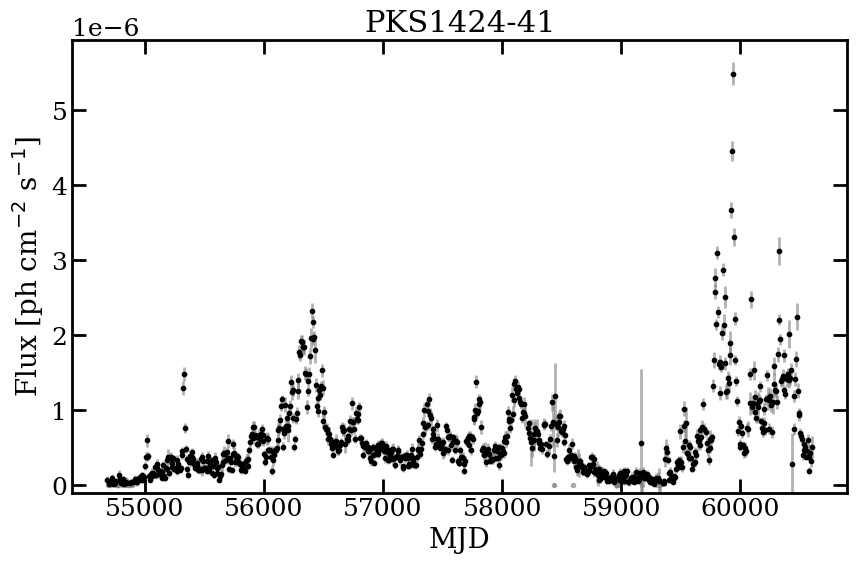}\vspace{-0.2cm}
\small
\caption{
    Light curves of the BL Lac object \mkn (\textit{left}) and the FSRQ \pks (\textit{right}) covering a time range from August 2008 until October 2024. All uncertainties and bins with TS$<1$ are plotted in grey.
    }
\label{fig:lcs}
\end{figure*}

\begin{figure*}
\centering
\includegraphics[width=0.47\textwidth]{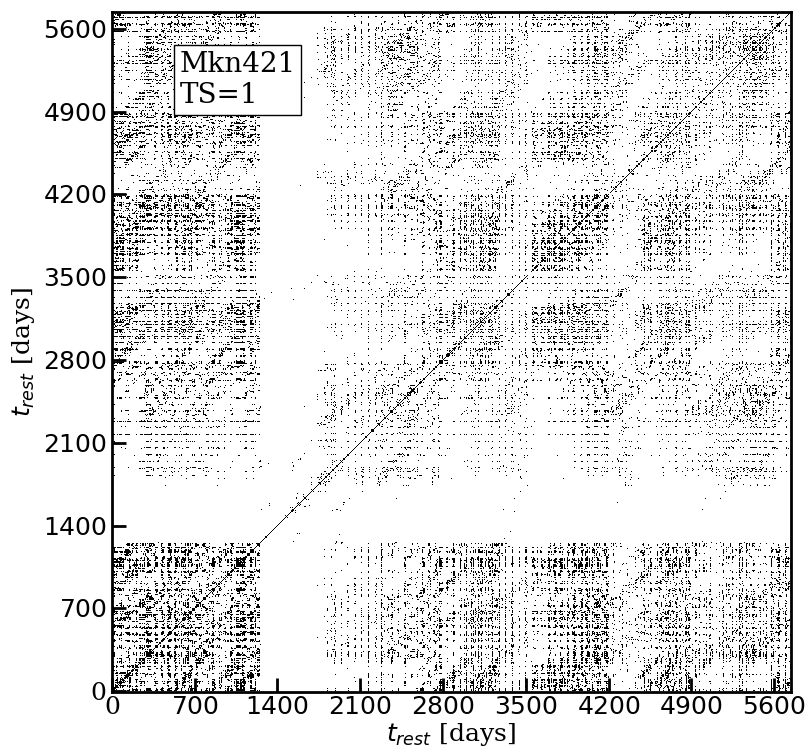} \hfill
\includegraphics[width=0.47\textwidth]{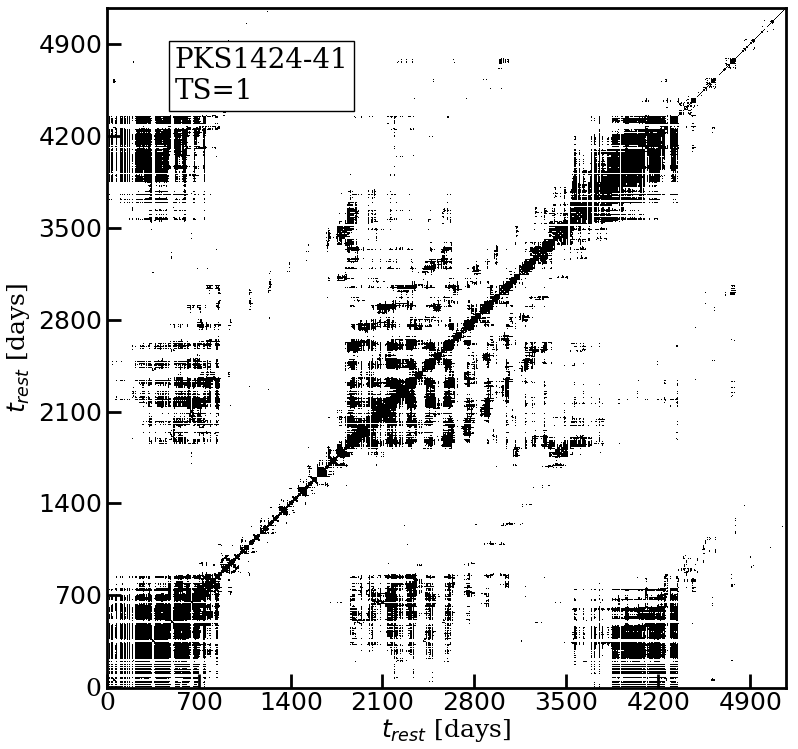}\vspace{-0.2cm}
\small
\caption{
Recurrence plots of \mkn (\textit{left}) and \pks (\textit{right}) derived from the lightcurves shown in Fig.~\ref{fig:lcs}. For both RPs, ten percent of the matrix is filled, meaning our choice of recurrence rate is 0.1.}
\label{fig:rps}
\end{figure*}

\section{Method}\label{sec:methods}

RPs were introduced by \cite{Eckmann1987} as a generalized means to visualize the recurrent phenomena of dynamical systems (see \citealt{Marwan2007} for a review). Suppose we have a dynamical system represented by the trajectory ${x_i}$ for $i=1,...,N$ in a $d$-dimensional phase space. In this case, ``phase space" refers to state space; for dynamical systems, each dimension, $n$, in state space typically corresponds to the $n^{th}$ derivative of the equations of motion, e.g., $(x, \dot{x}, \ddot{x}, ...)$. The rows and columns of the recurrence matrix correspond to pairs of points in time. We insert `1' in the matrix where two points in time are close in position in phase space (within a \textit{threshold}, $\epsilon$, neighborhood) and `0' otherwise. An RP is the graphical representation of the recurrence matrix. Critically, phase space can be reconstructed for scalar time series such as light curves (e.g., as shown in \citealt{Phillipson2020}), the most common being the Time Delay Method \citep{Sauer1991}, used in this work.

RPs result in square images with patterns unique to different dynamical systems. In fact, there are many summary statistics that describe an RP. In particular, some measures based on the diagonal-line features in the RP are mathematically equivalent to a variety of dynamical invariants underlying the observed time series \citep{Thiel2003}. Furthermore, RPs preserve all dynamical information of a system \citep{Robinson2009}. RPs and their summary statistics are widely used to distinguish deterministic from stochastic variability and detect quasi-periodicity, nonlinearity, and chaos (e.g., \citealt{Phillipson2023, Marwan2002}). 

We consider two summary statistics from the RP. Determinism (\textit{DET}) is defined as the ratio of recurrence points that form diagonal structures (at least two connected points in the RP in the diagonal direction) to all recurrence points and \textit{Lmax} is defined as the length of the longest diagonal line. \textit{DET} is a measure of the predictability of the system while \textit{Lmax} is related to the largest positive Lyapunov exponent \citep{Eckmann1987}, thus frequently used as a proxy for detecting nonlinearity in a system. The \textit{recurrence rate} is the percentage of recurrence points (non-zero entries) in the RP and is directly correlated to the threshold, as an increase of $\epsilon$ results in a higher-density recurrence matrix. 

\begin{figure*}[t]
\centering
\includegraphics[width=0.495\textwidth]
{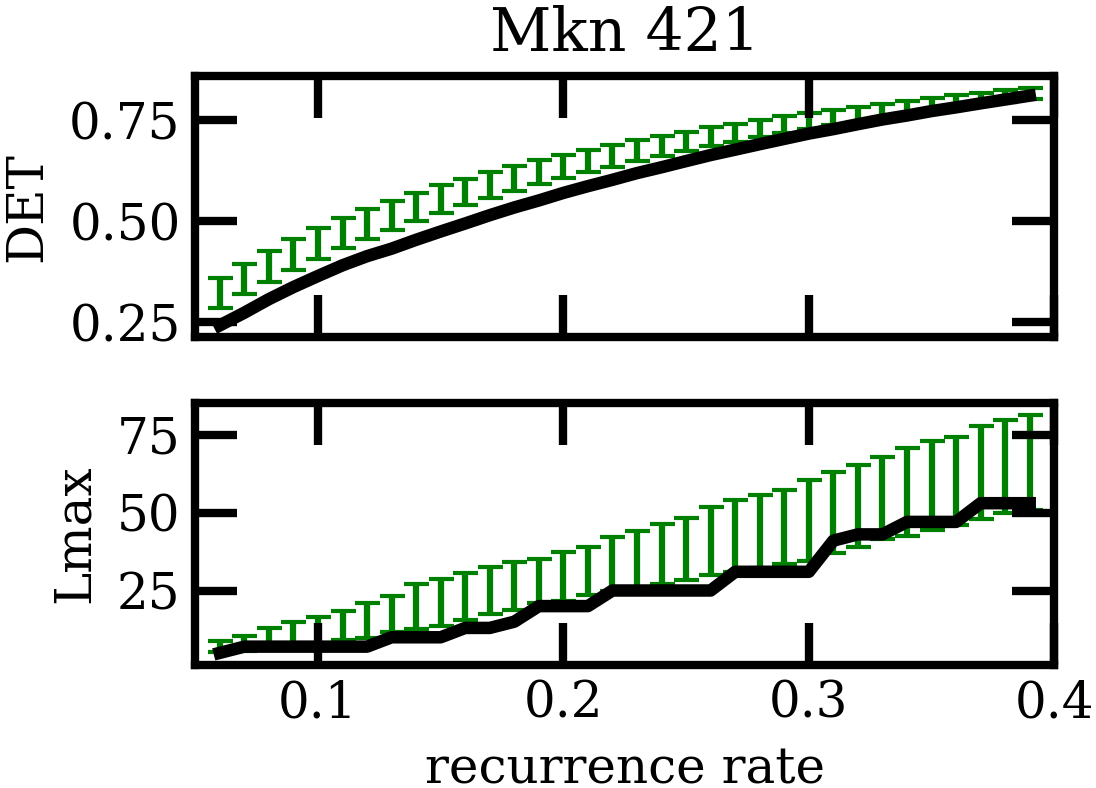}
\includegraphics[width=0.495\textwidth]
{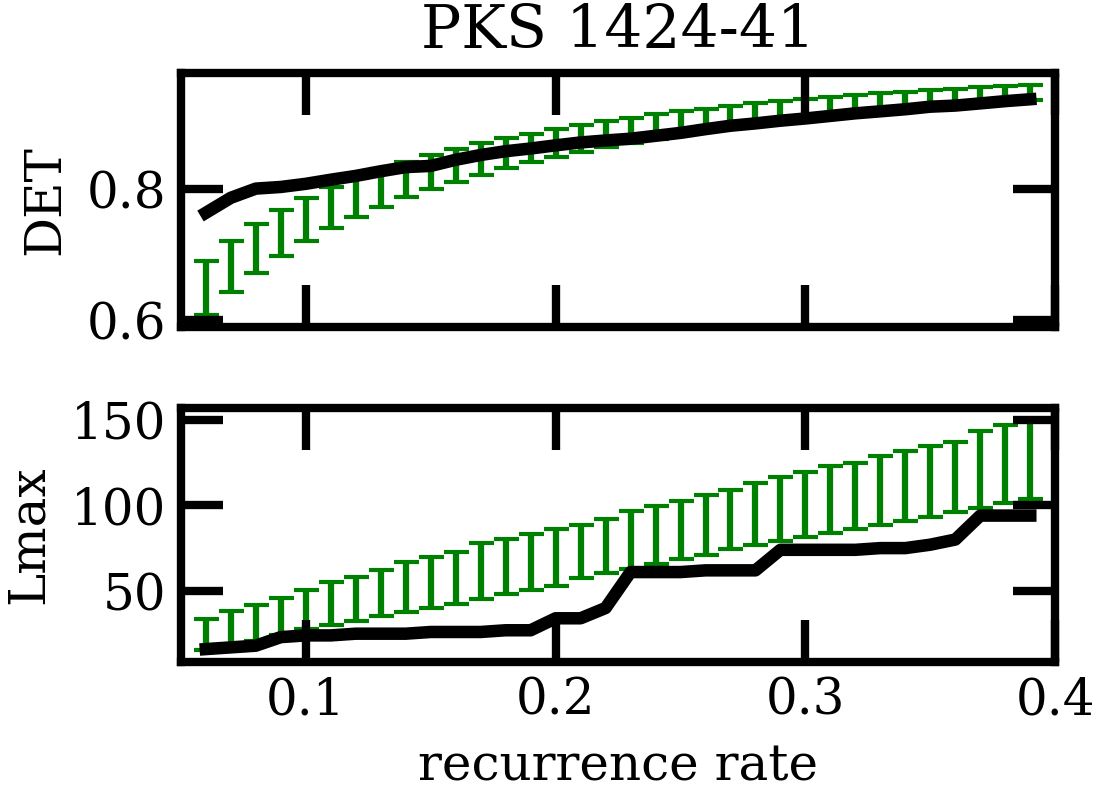}\vspace{-0.2cm}
\small
\caption{
Two RP measures as a function of \textit{recurrence rate} for \mkn (left) and \pks (right). The black line corresponds to the RP measure for the source light curve, while the green vertical bars represent the range of values for the ensemble of surrogates. \textbf{Top row}: The \textit{DET} measure, which probes the determinism and predictability of the source. \textbf{Bottom row}: The \textit{Lmax} measure, which is a proxy for nonlinearity in the source.}
\label{fig:recurrence}
\end{figure*}

Surrogate data testing \citep{Theiler1992} is used to determine the significance of nonlinearity, determinism, and other features in a dataset. It requires generating an ensemble of time series (called surrogates) that have identical statistical characteristics (e.g., autocorrelations) to the original time series but lack dynamical information (e.g., determinism), and then comparing the feature(s) of interest between the original time series and its surrogates. The surrogates are generated directly from the original time series itself (a constrained realizations approach) via a type of random process and therefore represent a \textit{null hypothesis} for the origination of the feature(s) of interest. A significant deviation of the original time series feature(s) from that of the surrogates supports a rejection of the null hypothesis. Here, we use Iterative Amplitude Adjusted Fourier Transform (IAAFT) surrogates \citep{Schreiber1996}, which preserve the power spectrum and flux distribution of the original time series. These correspond to a null hypothesis of the data representing a stationary linear Gaussian process, measured through an invertible, time-independent instantaneous measurement function.

\section{Results \& Discussion}\label{sec:results}

In this work, we present our results for the BL Lac object \mkn and the FSRQ \pks. We create 7-day binned \gm-ray light curves (see Fig.~\ref{fig:lcs}) from the public \lat data using the standard processing and fitting methods \cite[see, e.g., the detailed description in][]{gokus2024}. For our analysis, we consider all bins with a test statistic value of $\geq1$ (i.e., a detection filter including only data with $\geq1\sigma$). This excludes 1.9\% and 3.9\% of all bins in the light curves of \mkn and \pks, respectively. Note that besides the low threshold of $\sigma\geq1$, the large majority of bins in the light curves are detected above $5\sigma$ (96\% for \mkn and 81\% for \pks).

We produce RPs using the Time Delay Method, where the phase space embedding for each source can be constructed with a delay of 8 bins and 21 bins and a dimension of 4 and 6 for \mkn and \pks, respectively. 
The resulting RPs, which are shown in Fig.~\ref{fig:rps}, exhibit different characteristics. For \mkn, similar states of the light curve are repeated regularly and spread over the entire RP, creating a somewhat checkered pattern overlaid with fine diagonal structures throughout, indicating more stationary variability and recurrences. In contrast, for \pks the time range in which points of the light curve get close in phase space results in larger features overall but sparsely distributed, indicating stronger non-stationarity.

For validation, we compare both the \textit{DET} and \textit{Lmax} statistics with IAAFT surrogate light curves to determine the significance of the derived features. Evaluating an original time series against its surrogates for a range of thresholds, $\epsilon$, (and thus, recurrence rates) increases the validity of conclusions compared to selecting a single, fixed value for $\epsilon$ \citep{Zbilut2002}. In Fig.~\ref{fig:recurrence}, we investigate how much the \textit{DET} and \textit{Lmax} values differ from those computed for our set of surrogates for an increasing recurrence rate. 

We find that for a range of recurrence rates, \mkn shows a significant difference from its surrogates for \textit{DET}, while for \pks this is only observed for a small range at very low recurrence rates. This suggests that \mkn is the more deterministic one of the two sources.
For the \textit{Lmax} measure, both sources show significantly different values compared to their surrogates, which often suggests the presence of nonlinearity.
Detections of nonlinearity, however, are typically degenerate with nonstationarity \citep{Kugiumtzis2001}, and therefore it is recommended to perform complementary tests to facilitate distinguishing true nonlinearity from nonstationarity \cite[e.g.,][]{Mannattil2016,Phillipson2023}.
As we expect nonstationarity to be a prevalent issue for the majority of blazar light curves, we find that `vanilla' IAAFT surrogates might not be the ideal choice for a comparison; since they do not preserve nonstationarity \citep{Borgnat2009}, we cannot differentiate nonlinearity and nonstationarity from the \textit{Lmax} measure alone.
For future work, we will mitigate this by using more sophisticated versions of IAAFT surrogates that are capable of preserving the nonstationarity of the original time series, such as Pinned Wavelet IAAFTs \cite[PWIAAFTs;][]{Keylock2007}.
In addition, we will extend our analysis towards a larger sample of \gm-ray bright blazars ($\sim40-50$ sources) consisting of both FSRQs and BL Lac objects. In addition to probing each source for the dynamics in the blazar jets and QPOs on time scales of weeks to years, we will compare the derived RP statistics in context of the two main source classes.


\begin{acknowledgements}
The authors acknowledge support for this work via NASA grant 80NSSC25K7104.
\end{acknowledgements}

\bibliographystyle{aa}
\bibliography{bibliography}

\end{document}